
%
%
%
%
%
%
%
%
\input harvmac
\def\nl{\hfill\break}
\def\bar{\overline}
\def\lam{\lambda}
\def\ep{\epsilon}
\def\epp{\epsilon^{kink}}
\def\eep{\lam e^{-\ep}}
\def\eepp{\lam e^{-\ep(\tht')}}
\def\eept{\lam e^{-\ep(\tht)}}
\def\tht{\theta}
\def\infinity{\infty}
\def\piR{\left({\pi\over 6 R}\right)}

\nref\rDMB{R. Dashen, S.K. Ma and H.J. Bernstein, Phys. Rev. 187 (1969) 345.}
\nref\rZandZ{A.B. Zamolodchikov and Al.B. Zamolodchikov, Ann. Phys. 120
(1980) 253.}
\nref\rYY{C.N. Yang and C.P. Yang, J. Math. Phys. 10 (1969) 1115.}
\nref\rAZi{A.B. Zamolodchikov, Adv. Stud. Pure Math. 19 (1989) 1.}
\nref\rMS{G. Mussardo and G. Sotkov, submitted to Phys. Rep. C.}
\nref\rAlZi{Al.B. Zamolodchikov, Nucl. Phys. B342 (1990) 695.}
\nref\rMari{M.J. Martins, Phys. Lett. 257B (1991) 317.}
\nref\rMar{M.J. Martins, Phys. Rev. Lett. 67 (1991) 419.}
\nref\rKMiv{T. Klassen and E. Melzer, ``Spectral Flow between
Conformal Field Theories in 1+1 Dimensions'', Chicago preprint EFI 91-17,
Miami preprint UMTG-162.}
\nref\rCarBC{J.L. Cardy, Nucl. Phys. B275 (1986) 200.}
\nref\rCarVer{J.L. Cardy, Nucl. Phys. B324 (1989) 581.}
\nref\roldstuff{M. Gaudin, Phys. Rev. Lett. 26 (1971) 1305;\nl
M. Takahashi and M. Suzuki, Prog. Th. Phys. 48 (1972) 2187;\nl
 J.D. Johnson and B. McCoy, Phys. Rev. A6 (1972) 1613; \nl
A.M. Tsvelick and P.B. Weigmann, Adv. Phys. 32 (1983) 453.}
\nref\rKR{A.N. Kirillov and N. Yu. Reshetikhin, J. Phys. A20 (1987) 1587.}
\nref\rTsv{A.M. Tsvelik, Sov. J. Nucl. Phys. 47 (1988) 172.}
\nref\rKMi{T.R. Klassen and E. Melzer, Nucl. Phys. B338 (1990) 485.}
\nref\rKMii{T.R. Klassen and E. Melzer, Nucl. Phys. B350 (1990) 635.}
\nref\rNon{M.J. Martins, Phys. Lett. 240B (1990) 404;\nl
P. Christe and M.J. Martins, Mod. Phys. Lett. A5 (1990) 2189.}
\nref\rIM{H. Itoyama and P. Moxhay, Phys. Rev. Lett. 65 (1990) 2102.}
\nref\rAlZii{Al.B. Zamolodchikov, Nucl. Phys. B358 (1991) 497.}
\nref\rAlZiii{Al.B. Zamolodchikov, Nucl. Phys. B358 (1991) 524.}
\nref\rAlZiv{Al.B. Zamolodchikov, ``Resonance Factorized Scattering and
Roaming Trajectories'', Ecole Normale preprint ENS-LPS-355.}
\nref\rFI{P. Fendley and K. Intriligator, ``Scattering and Thermodynamics
of Supersymmetric Solitons'', Boston preprint BUHEP-91-17, Harvard preprint
HUTP-91/A043.}
\nref\rLassig{M. L\"assig, ``Exact Universal Amplitude Ratios in
Two-Dimensional Systems Near Criticality'', Santa Barbara preprint
UCSBTH 91-20.}
\nref\rCarMod{J.L. Cardy, Nucl. Phys. B270 (1986) 186.}
\nref\rEVer{E. Verlinde, Nucl. Phys. B300 (1988) 360;\nl
 G. Moore and N. Seiberg, Phys. Lett. 212B (1988) 451.}
\nref\rBCN{H.W.J. Bl\"ote, J.L. Cardy and M.P. Nightingale, Phys. Rev.
Lett. 56 (1986)742;\nl I. Affleck, Phys. Rev. Lett., 56 (1986) 746.}
\nref\rYang{S.-K. Yang, Nucl. Phys. B285 (1987) 639.}
\nref\rLewin{L. Lewin, {\it Polylogarithms and Associated Functions}
(North-Holland, 1981).}
\nref\rZamii{A.B. Zamolodchikov, Int. J. Mod. Phys. A3 (1988) 743.}
\nref\rZF{A.B. Zamolodchikov and V.A. Fateev, Sov. Phys. JETP 62 (1985)
215;\nl JETP 63 (1986) 913.}
\nref\rKS{R. Koberle and J.A. Sweica, Phys. Lett. 86B (1979) 209.}
\nref\rEYHM{T. Eguchi and S.-K. Yang, Phys. Lett. 224B (1989) 373;\nl T.
Hollowood and P. Mansfield, Phys. Lett. 226B (1989) 73.}
\nref\rAlZv{Al. B. Zamolodchikov, Phys. Lett. 253B (1991) 391.}
\nref\rQiu{Z. Qiu, Nucl. Phys. B295 (1988) 171.}
\nref\rGin{P. Ginsparg, Nucl. Phys. B295 (1988) 153.}
\nref\rBCDS{H.W. Braden, E. Corrigan, P.Dorey and R. Sasaki, Phys. Lett.
227B (1989) 411; Nucl. Phys. B338 (1990) 689.}
\nref\rCMFKM{P. Christe and G. Mussardo, Nucl. Phys. B330 (1990) 465;\nl
P. Freund, T.R. Klassen and E. Melzer, Phys. Lett. 229B (1989) 243.}
\nref\rFZSZ{V.A. Fateev and A.B. Zamolodchikov, Int. J. Mod. Phys. A5
(1990) 1025;\nl G. Sotkov and C.-J. Zhu, Phys. Lett. 229B (1989) 391.}
\nref\rZamkink{A.B. Zamolodchikov, Sep. 1989 preprint.}
\nref\rssG{I.Ya. Arefyeva and V.E. Korepin, JETP Lett. 20 (1974) 680; \nl
S.N Vergeles and V.M. Gryanik, Sov. J. Nucl. Phys. 23 (1976) 1324; \nl
B. Schroer, T.T Truong and P.H. Weisz, Phys. Lett. 63B (1976) 422.}
\nref\rLC{A. Ludwig and J.L. Cardy, Nucl. Phys. B285 (1987) 687.}
\nref\rZamLG{A.B. Zamolodchikov, Sov. J. Nucl. Phys. 46 (1987) 1090.}
\nref\rKMiii{T.R. Klassen and E. Melzer, ``On the Relation between Scattering
Amplitudes and Finite-Size Mass Corrections in QFT'', Chicago preprint EFI
90-79, Miami preprint UMTG-160.}
\nref\rLM{M. L\"assig and M.J. Martins, Nucl. Phys. 354B (1991) 666.}
\nref\rDJMO{E. Date, M. Jimbo, T. Miwa and M. Okado, Phys. Rev. B35 (1987)
2105.}

\Title{\vbox{\baselineskip12pt\hbox{BUHEP-91-16}\hbox{}
                \hbox{}}}
{\vbox{\centerline{Excited-state thermodynamics}}}

\centerline{Paul Fendley}
\bigskip\centerline{Department of Physics}
\centerline{Boston University}\centerline{590 Commonwealth Avenue}
\centerline{Boston, MA 02215}
\vskip 1.0cm

In the last several years, the Casimir energy for a variety of 1+1-dimensional
integrable models has been determined from the exact S-matrix. It is shown
here how to modify the boundary conditions to project out the lowest-energy
state, which enables one to find excited-state energies. This is done by
calculating thermodynamic expectation values of operators which generate
discrete symmetries.  This is demonstrated with a number of perturbed
conformal field theories, including the Ising model, the three-state Potts
model, ${\bf Z}_n$ parafermions, Toda minimal S-matrices, and massless
Goldstinos.

\Date{August 1991}

\vfill\eject

\newsec{Introduction}

A wealth of physical information about a relativistic quantum particle theory
is contained in the knowledge of the asymptotic particle states and their
scattering matrix. This includes all thermodynamic quantities of an
infinite-volume gas of these particles, as long as the gas is dilute enough so
that a particle description is valid. There is a closed-form expression for
the virial coefficients (terms in the expansion of the free energy in the
fugacity) depending only on the S-matrix\rDMB. Doing the calculation in this
manner is usually unwieldy and unnecessary, because one can generally
calculate thermodynamic quantities directly from perturbative field theory.
However, the situation is markedly different in 1+1-dimensional integrable
theories. Here the S-matrix has a number of properties which simplify matters
a great deal: it is completely elastic (momenta are conserved individually in
a collision), the $n$-body S-matrix factorizes into a product of two-body
ones, and one often can derive or conjecture it exactly\rZandZ.  It is in
principle possible to calculate these quantities using perturbed conformal
field theory, but using the S-matrix is often more tractable, and if the
S-matrix is exact, the results are non-perturbative.

The techniques for calculating the Casimir energy (the lowest eigenvalue of
the one-dimensional quantum Hamiltonian) were first used in the
non-relativistic case in \rYY. Recently, there has been a great deal of
interest in this field, because the study of perturbed conformal field theory
has enabled conjectures for a large number of exact S-matrices\rAZi.\foot{For
a review of much of this activity, see \rMS.} The calculation of the Casimir
energy in such a case goes by the name of the Thermodynamic Bethe Ansatz (TBA
for short) and was first done for the Yang-Lee model and the three-state Potts
model in ref.\rAlZi. This paper will discuss modifications of these techniques
recently proposed in \rMari--\rKMiv, which enable the calculation of energies
of excited states.

The result of the TBA is a set of coupled integral equations, one for each
particle in the spectrum. These equations give the exact Casimir energy as a
function of the volume of space. Although one can solve these equations only
numerically, it is possible to extract either the infrared or ultraviolet
asymptotics. The ultraviolet limit is where the theory becomes a conformal
field theory, and thus the results can be compared with results at the
conformal point. The results of this paper will be compared to conformal field
theories with a variety of boundary conditions\rCarBC\rCarVer.
The TBA equations have been studied in a large variety of exactly solvable
theories, including the XXZ spin chain in a magnetic field\roldstuff\ and its
generalizations\rKR, the supersymmetric sine-Gordon model\rTsv, all the
simply-laced affine Toda theories\rKMi\rKMii, a number of non-unitary
models\rAlZi\rNon, the perturbed minimal models\rIM--\rAlZiv, and
the perturbed N=1 and N=2 supersymmetric minimal models\rFI. Some of these
results may even be experimentally verifiable\rLassig.

The TBA allows the calculation of thermodynamic quantities for a
one-dimensional system, where the spatial dimension is a circle of
circumference $L$, with $L\rightarrow\infinity$. Any thermodynamic expectation
value of a $d$-dimensional quantum system at temperature $T$ can be
alternatively viewed as an expectation value in a $(d+1)$-dimensional
Euclidean quantum field theory, where the additional dimension is a circle of
circumference $R=1/T$. This is therefore equivalent to a two-dimensional
theory with spacetime a torus of periodicity ($R,L$).  We can view either the
$R$-direction or the $L$-direction as space, with the other direction then
taking the role of Euclidean time.  Thus the partition function $Z(R,L)$ is
equal to both
\eqn\ZRLi{Z(R,L) = {\hbox{tr}}\left[ e^{-RH_L}\right] ,}
and
\eqn\ZRLii{Z(R,L) = {\hbox{tr}}\left[ e^{-LH_R}\right] ,}
where $H_C$ is the Hamiltonian for the system with space a circle of
circumference $C$. This equivalence has proven very useful in conformal field
theory defined on a torus, where it goes by the name of modular invariance,
and puts severe constraints on the operators allowed in a
theory\rCarMod\ and on their fusion rules\rEVer.
In the $L\rightarrow\infinity$ limit, the only
state which contributes to the partition function is the ground state.
Thus
\eqn\ZRL{Z(R,L\rightarrow\infty)\longrightarrow e^{-E(R)L},}
where $E(R)$ is the ground-state energy (lowest eigenvalue of $H_R$)
of the theory on a circle of circumference $R$, and does not depend on $L$.
In any thermodynamic system the free energy is $F=-\ln{Z}/R$,
so \ZRL\ yields
\eqn\FRL{E(R)={R\over L}F(R,L\rightarrow\infty).}
Therefore, the Casimir energy can be determined from the TBA.

An important check on these results can obtained by studying the limit where
the particle masses go to zero, so the system approaches a conformal field
theory. A general result of conformal field theory \rBCN\ predicts
\eqn\BCN{E(R) ={ 2\pi\over R}\left( h+\bar h -{c\over 12} \right), }
where $h$ and $\bar h$ are respectively the left and right conformal
dimensions of the operator which creates the lowest energy state when acting
on the vacuum. For a unitary theory with periodic boundary conditions
(anti-periodic for any fermions present), this operator is the identity, so
that $h=\bar h=0$. This relation gives a very useful way of checking that the
conjectured S-matrices used in the TBA calculation are consistent---\BCN\
relates a TBA result in the $m\rightarrow 0$ limit to results obtainable from
conformal field theory.
This check has been successfully done in many cases\rKMi--\rFI, thus
giving even stronger support for the conjectured S-matrices of these models.
In addition, many of these papers numerically
calculate corrections to \BCN, and compare them with results obtained with
perturbation theory around the conformal point.

All of this work determined $E(R)$ for the lowest-energy state of the theory.
However, the TBA formalism is not limited to this type of calculation.  In
several recent papers the TBA equations for a number of models were modified
in order to describe the behavior of excited-state energies\rMari--\rKMiv.  To
check these TBA equations, the values of $h$ in \BCN\ were calculated, and
were found to match those of particular operators in the conformal theory. The
equations were solved numerically in some cases, and the results agreed with
the appropriate perturbed conformal field theory.

In this paper we demonstrate two methods of deriving excited-state energies.
These methods are general and allow the determination of excited-state
energies in any model with a discrete symmetry. This relates the results of
\rMar\ and \rKMiv\ for massive theories to conformal results\rCarBC, and gives
a simple physical interpretation of their modified TBA equations.

Both methods involve the calculation of the thermodynamic expectation value of
the operators which implement this discrete symmetry, and are based on the way
partition functions are calculated in conformal field theories with boundary
conditions\rCarBC\rCarVer.  The unnormalized expectation value of an operator
$A$ is
\eqn\Ai{<A>={\hbox{tr}}\left[Ae^{-H_L R}\right].}
Thus $Z(R,L)=<1>$. Just as $<1>$ can be rewritten as \ZRLii\ by interchanging
space and Euclidean time, we rewrite
\eqn\Aii{<A>={\hbox{tr}}\left[e^{-\tilde H_A L}\right],}
where $\tilde H_A$ is the Hamiltonian for the same model, but with new
boundary conditions depending on the operator $A$. If $A=1$, then the boundary
conditions are periodic for bosons and antiperiodic for fermions, as is
standard in thermodynamics. Another simple example is the case
of a free scalar field $\phi$. When $A$ is the operator which sends
$\phi\rightarrow -\phi$, $\tilde H_A$ is the Hamiltonian for the model with
antiperiodic boundary conditions.

As in the periodic case, in the $L\rightarrow\infinity$ limit
\eqn\EAR{E_A(R)={R\over L}\ln{<A>} ,}
where $E_A(R)$ is the lowest eigenvalue of $\tilde H_A$. In the $m\rightarrow
0$ limit, \BCN\ is recovered as long as the energy-momentum tensor is invariant
under the action of $A$. However, $h$ is not necessarily zero, even in a
unitary theory. This is because the boundary conditions can project the
identity state out of the spectrum. In conformal field theory, this has been
demonstrated explicitly for the Ising model and the three-state Potts model
with a variety of boundary conditions\rCarBC.

There are two ways of calculating $<A>$ when $L\rightarrow\infty$, thus
determining $E_A(R)$.  The first method utilizes an imaginary chemical
potential and is used when $A$ is diagonal on the space of particles. The
second method works when $A$ interchanges particles, and involves truncating
the Hilbert space. These methods can be combined for a general $A$.

Section 2 is a warm-up for the rest of the paper---both methods of calculating
excited-state energies are applied to the Ising model, where the TBA is not
needed.  In section 3, the TBA in the presence of a general chemical potential
is reviewed, and it is shown how this can be used to calculate expectation
values of diagonal operators. Section 4 demonstrates within the TBA how one
truncates the Hilbert space to obtain expectation values of non-diagonal
operators. Section 5 applies these methods to models described by simply-laced
Toda minimal S-matrices, including the three-state Potts model. Section 6
discusses two more models, the massless perturbation of the tricritical Ising
model and a model which displays many of the characteristics of the minimal
models.  Section 7 contains conclusions and outlines possible future work.

\newsec{The Ising Model}

The field theory describing the $T\ne T_c$ Ising model in two spacetime
dimensions consists of a single free Majorana fermion of mass
$m$.\foot{Another field theory describing the Ising model consists of a single
massive boson with S-matrix $S=-1$.  Which of the two is applicable depends on
the boundary conditions, since they have different finite-volume
thermodynamics\rKMii. However, the ground-state energy is the same in both
theories, so it does not matter which theory we use here.} We consider an
ensemble of $N$ particles, and the rapdity $\tht_i$ of the $i$th particle is
defined by
\eqn\rap{\eqalign{p_i&=m\sinh\tht_i\cr E_i&=m\cosh\tht_i.\cr}}
Space is taken to be periodic with period $L$, so each momentum is quantized:
\eqn\pquant{p_i={2n_i\pi\over L}.}
When the particle number $N$ is large,
the partition function
at temperature $T=1/R$ and chemical potential $\mu$ is defined as
\eqn\Zferm{<1>_\lam =Z_\lam(R,L)=\tr\left[\lam^N e^{-RH_L}\right],}
where the fugacity $\lam\equiv e^{\mu R}$.  This is the standard
one-dimensional free-fermion partition function, which is
\eqn\ZIsing{Z_\lam(R,L)=\prod_{n=-\infty}^{+\infty}(1+\lam e^{-RE_n}),}
where
\eqn\forEn{E_n=\left(({2\pi\over L}n)^2+m^2\right)^{1/2}.}
Thus
\eqn\Fdiscrete{F_\lam\equiv -T\ln Z_\lam=-T\sum_{n=-\infty}^{\infinity}
\ln (1+\lam e^{-RE_n}).}
In the limit $L\rightarrow\infty$, we can replace the sum over $n$ with an
integral, and using
$$dn={L\over 2\pi}dp={L\over 2\pi}m\cosh\tht d\tht$$
yields
\eqn\Fferm{F_\lam=-{mL\over 2\pi R}\int \cosh\tht \ln(1+\eept) d\tht,}
where $\ep(\tht)\equiv mR\cosh\tht$ is the one-particle energy.

By definition, the eigenvalues of diagonal operators vary only with particle
number. Looking at \Zferm, one sees that to calculate the expectation value of
such operators, one needs only to modify the chemical potential appropriately.
Thus
\eqn\minone{<(-1)^N >=<1>_{\lam=-1},}
This corresponds to a chemical potential of $i\pi T$. Even though this is
imaginary, \Fferm\ is perfectly well-defined by analytic continuation in
$\lam$.  In fact, $F_{-1}$ is minus the free energy of a free boson.  The
argument of \Fferm\ has a singularity at $\lam=-1$ and $\tht=0$, but the
integral is well-defined.

All of this analysis has been done treating the $L$-direction as space. When
$R$ is the space dimension, we recover the picture of
\Aii. When $A=1$, $\tilde H_A$ is the Hamiltonian for free fermions with
antiperiodic boundary conditions. When $A=(-1)^N$, the boundary conditions in
the $R$-direction are flipped: $\tilde H_A$ is the Hamiltonian for free
fermions with periodic boundary conditions. In the language of Ising spins,
these correspond to periodic and antiperiodic boundary conditions on the
spins, respectively. Thus with the equivalence \EAR, evaluating $<(-1)^N>$
gives the ground-state energy for the Ising model with antiperiodic boundary
conditions.

In the limit $m\rightarrow 0$, this model approaches the Ising conformal field
theory. To evaluate $\Fferm$ in this limit involves a few tricks \rKR\rAlZi\
which are discussed in section 3. Using the definition
\EAR\ of the ground-state energy $E_A(R)$, the result is
\eqn\EIsing{\eqalign{&E_1(R)=-{1\over 2} \left({\pi\over 6R}\right)\cr
&E_{(-1)^N}(R)=\piR\cr }}
The second equation should be clear even without the tricks since it is minus
the energy of a free boson, which has $c=1$. Equation \BCN\ yields the scaling
dimensions of the operator which creates these two states. $E_1(R)$ yields
$c=1/2$, because $h=\bar h=0$. $E_{(-1)^N}(R)$ shows that the operator with
conformal dimensions $(1/16,1/16)$ creates the lowest-energy state in the
Ising model with antiperiodic boundary conditions. This operator is the spin
field $\sigma$.  This result, of course, can be derived at the conformal
point\rCarBC.

We turn to operators which interchange particles. Since there is only one
particle in the Ising model, this is not relevant here. However, in a system
consisting of two decoupled Ising models, there is an operator which exchanges
a fermion of the first system $\psi_I(\tht)$ with one from the second system
$\psi_{II}(\tht)$. This operator, which we denote by $K$, commutes with the
Hamiltonian, so we can compute $<K>$.  The Hamiltonian is diagonal on the
space of states in theories with diagonal S-matrices, and we take the states
to be orthonormal. Therefore, a state must be invariant under $K$ to give a
non-zero contribution to the trace in $<K>$.  We define the truncated Hilbert
space ${\cal H}_K$ as the space of such states; i.e., states where for every
$\psi_I$ with a rapidity $\tht$, there is also a $\psi_{II}$ with the same
rapidity $\tht$.  Then
\eqn\forKlam{\eqalign{<K>_\lam=&\tr_{{\cal H}_K} \left[
\lam_I^{N/2}\lam_{II}^{N/2} e^{-RH_L}\right]\cr
=&\sum_{{\cal H}_K} \lam_I^{N/2}\lam_{II}^{N/2} e^{-E_K},\cr}}
where $$E_K=m\sum_{i=1}^{N}\cosh(\tht_i)=2m\sum_{i=1}^{N/2}\cosh(\tht_i).$$
The configuration sum is that of a single Ising model with $N/2$ particles,
and the energy is that of a single Ising model of mass $2m$. Thus
\eqn\KIsing{<K>_\lam=\prod_n(1+\lam_I\lam_{II} e^{-2E_n}),}
where $E_n$ is defined in \forEn. In the $L\rightarrow\infty$ limit,
\eqn\Kferm{\ln <K>_\lam=-{mL\over 2\pi R}\int \cosh\tht \ln(1+\lam_I\lam_{II}
e^{-2\ep(\tht)}) d\tht,}
where $\ep(\tht)=mR\cosh(\tht)$ is still the one-particle energy. Notice that
in \Kferm\ there is no extra factor of $2$ in front of the integral.

As before, we interchange space and Euclidean time, and interpret $E_K$ as the
ground-state energies for models with different boundary conditions. In the
language of Ising spins, $\tilde H_K$ for $\lam_I=\lam_{II}=1$ is the
Hamiltonian for two Ising models, coupled only by the boundary condition
$\sigma_I(0,y)=\sigma_{II}(R,y)$ and $\sigma_I(R,y)=\sigma_{II}(0,y)$, where
$\sigma_I(x,y)$ and $\sigma_{II}(x,y)$ are the Ising spins at site $(x,y)$.
When $\lam_I=-1$ and $\lam_{II}=1$, this amounts to calculating
$<K(-1)^{N/2}>$.
(Remember, in the truncated Hilbert space $N/2$ is an integer.) This
corresponds to the boundary condition $\sigma_I(0,y)=-\sigma_{II}(R,y)$ and
$\sigma_I(R,y)=\sigma_{II}(0,y)$.

In the conformal limit $m\rightarrow 0$, \Kferm\ is independent of $m$.
Comparing this with \Fferm, we see that
\eqn\Klim{\lim_{m\rightarrow 0} \left(\ln <K>_\lam\right)=
{1\over 2} \lim_{m\rightarrow 0} F_\lam.}
Thus we can read off the answers from \EIsing, and when $\lam=1$,
\eqn\EKi{E_K(R)=-{1\over 4} \piR.}
Since we have two Ising models, $c=1$, and $h+\bar h=(1-1/4)/12$. Thus the
field which creates this state has conformal dimensions $(1/32,1/32)$. When
$\lam=-1$,
\eqn\EKii{E_K(R)=+{1\over 2}\piR .}
Here $h+\bar h=(1+1/2)/12$, and the conformal dimensions are $(1/16,1/16)$.
Both of these results are in agreement with those found using modular
transformations at the conformal point\rYang.

\newsec{The TBA with a chemical potential}

In this section we briefly derive the TBA equations in the presence of a
chemical potential, following the treatment of refs.\rAlZi\rKMii.

The starting point of the TBA is the exact two-body elastic S-matrix.  For
simplicity, in this paper we will treat only diagonal S-matrices, although the
TBA can also be extended to models with non-diagonal S-matrices, where one
must introduce massless pseudo-particles into the TBA system of
equations\rAlZii.  The S-matrix element for the process
$a(\theta_1)b(\theta_2)\rightarrow b(\theta_2)a(\theta_1)$ is denoted by
$S_{ab}(\theta)$, where $\theta\equiv\theta_1-\theta_2$, the difference of the
particles' rapidities.  An S-matrix element in one spatial dimension is
defined so that when two particles $a(\theta_1)$ and $b(\theta_2)$ are
exchanged, the wavefunction is multiplied by $S_{ab}(\theta)$.

For the moment, we specialize to a theory with only one kind of particle, so
that there is only one S-matrix element $S(\tht)$.
The TBA applies to systems with a large number of particles $N$, where the
space
direction is periodic with
period $L$. The periodicity leads to a quantization condition on each
momentum $p_i=m\sinh(\theta_i)$:
\eqn\quant{e^{ip_iL}\prod_{j\ne i} S(\theta_i-\theta_j) = 1.}
Defining $\rho_r(\tht)$ to be the rapdity density (i.e.,
$\rho_r(\tht)\Delta\tht$ is the number of particles with rapidity between
$\tht$ and $\tht +\Delta\tht$), the logarithm of \quant\ is
\eqn\quantt{mL\sinh(\theta_i)-i\int
d\tht'\rho_r(\tht')\ln{S(\theta_i-\theta')} =2\pi n_i,}
where an integer $n_i$ is associated with each particle. All integrals in this
paper run from $-\infty$ to $+\infty$ unless otherwise labeled. This equation
is the generalization to an interacting model of the one-particle relation
$p_i= 2\pi n_i /L$. Here there is a set of coupled equations for the momenta.
We introduce a level density $\rho(\tht)$, so that $\rho(\tht)\Delta\tht$ is
the number of values of $\tht_i$ which solve
\quantt\ in the range $\tht + \Delta\tht$. Taking the large $L$ limit,
\quantt\ becomes
\eqn\forrho{\rho(\tht)= mL\cosh(\theta)+
\int d\tht'\rho_r(\tht')\phi(\theta-\theta'),}
where $$\phi(\tht)\equiv -i{\del\ln{S(\tht)}\over\del\tht}.$$

To determine the free energy, we use thermodynamics, as opposed to the
statistical mechanics used in section 2. In a system with chemical potential
$\mu$ ($\lam\equiv e^{\mu/T}$) at temperature $T$, we use
$F_\lam={\cal E}-TS-\mu N$ and minimize $F_\lam$ using \forrho\
as a constraint. Here the energy ${\cal E}$ is
\eqn\forH{{\cal E}=\int \rho_r(\tht)m\cosh\tht d\tht,}
the entropy $S$ is
\eqn\forS{S=\int d\tht[\rho\ln\rho -\rho_r\ln\rho_r -
(\rho-\rho_r)\ln(\rho-\rho_r)],}
and the particle number $N$ is
\eqn\forN{N=\int \rho_r (\tht) d\tht.}
In \forS, we have assumed that all particles are of ``fermionic'' type, with
at most one particle of a given species with a given momentum $\tht_i$. This
is true for all known one-dimensional theories, save a free boson.\foot{See
\rAlZi\ or \rKMi\
for a detailed explanation of this point. It is easy to treat the case where
particles are ``bosonic''---just replace $\rho$ in \forS\ with $(\rho +
\rho_r)$.} Defining the pseudoenergy $\ep(\tht)$ by\foot{This definition
includes the chemical potential, as opposed to \rKMii.}
\eqn\defep{\rho_r=\rho{\eep\over 1+\eep},}
the minimization of $F_\lam$ yields an integral equation for $\ep(\tht)$:
\eqn\forep{\ep(\tht)=mR\cosh\tht -{1\over 2\pi}\int
d\tht'\phi(\tht-\tht')\ln(1+\eepp).}
In terms of $\ep$, the free energy is
\eqn\fF{F_\lam=-{mL\over 2\pi R}\int d\tht \cosh\tht\ln(1+\eept).}
In the case where there is more than one species of particle, one introduces
particle and level densities for each species $a$, and finds
\eqn\bigone{\rho^a(\tht)= m_aL\cosh(\theta)+\sum_b \int
d\tht'\rho_r^b(\tht')\phi_{ab}(\theta-\theta'),}
where
\eqn\forphi{\phi_{ab}(\tht)\equiv -i{\del\ln{S_{ab}(\tht)}\over\del\tht}.}
The pseudoenergies are defined as in \defep, and minimizing the free energy
yields the integral equation
\eqn\forepa{\ep_a(\tht)=m_a R\cosh\tht -\sum_b{1\over
2\pi}\int
d\tht'\phi_{ab}(\tht-\tht')\ln(1+\lam_b e^{-\ep_b(\tht)}).}
We define the scaled free energy density $f_\lam(mR)$ as
$$f_\lam(mR)\equiv {R^2\over L} F_\lam,$$
and it is given by
\eqn\ffa{\eqalign{&f_\lam(mR)=\sum_a f_\lam^a(mR),\cr
&f_\lam^a(mR)\equiv -{mR\over 2\pi}\int d\tht \cosh\tht\ln(1+\lam_a
e^{-\ep_a(\tht)}).\cr}}

The conformal limit of \ffa\ is the limit $mR\rightarrow 0$. To evaluate this,
we define $\epp(\tht)\equiv\ep(\tht - x)$, where
$x=\ln(mR/2)$. When $x\rightarrow -\infty$, this yields equations
which do not depend on $mR$:
\eqn\epkink{\epp_a(\tht)= e^{\tht}-\sum_b{1\over
2\pi}\int
d\tht'\phi_{ab}(\tht-\tht')\ln(1+\lam_b e^{-\epp_b(\tht)}).}
and
\eqn\fcft{f_\lam^a(0)=-{1\over \pi}\int e^{\tht} \ln(1+\lam_a
e^{-\epp_a(\tht)}).}
There are a few tricks which enable us to evaluate this without knowing the
full function $\ep(\tht)$. We take the derivative of \epkink\ and substitute
for $e^\tht$ in \fcft. The $\phi_{ab}$ can be removed from the resulting
expression by using \epkink again. Every term depends on $\tht$ only through
$\epp$, and also has a $\del\epp/\del\tht$ in it. Since the shifted
$\ep(\tht)$ is always decreasing, the integral over $\tht$ can be replaced by
an integral over $\epp$, yielding
\eqn\fcftepa{2f_\lam(0)=-{1\over \pi} \sum_a\int_{C_a} d\ep\left[{\ep \lam_a
e^{-\ep} \over
1+\lam_a e^{-\ep}}+\ln(1+\lam_a e^{-\ep})\right],}
where the contour $C_a$ runs in the complex-$\ep$ plane from $\epp_a(-\infty)$
to $\epp_a(\infty)$. In theories without massless particles,
$\epp(\infty)=\infty$.  To determine $\epp(-\infty)$, we use the property of
$\epkink$ that $\epp(\tht)$ is flat from $\tht=-\infty$ to around $\tht=0$,
and that $\phi_{ab}$ falls off exponentially. This enables us to pull
$\ln(1+\lam_b e^{-\epp_b(\infty)})$ out of the integral in \forepa, which
yields
a system of equations for $x_a\equiv \exp(\epp_a(-\infty))=\exp(\ep_a(0))$:
\eqn\forxa{x_a=\prod_b(1+{\lam_b \over x_b})^{N_{ab}},}
where
\eqn\fornab{N_{ab}\equiv - {1\over 2\pi}\int\phi_{ab}(\tht) d\tht.}
When $\lam_a=1$, the contour $C_a$ lies on the real axis, but in the cases with
imaginary chemical potential we will discuss later, it may not. Of course,
one can deform $C_a$, paying attention to the logarithmic cuts in \fcftepa\
starting at $\mu_a-\ep_a=i(2n+1)\pi$.

We write \fcftepa\ in the form
\eqn\usingdilog{f_\lam(0)=-{1\over\pi}\sum_a {\cal L}_{\lam_a}(x_a),}
When $\lam_a$ are either $1$ or $-1$, and all $x_a\ge 1$ for $a$ where
$\lam_a=-1$ , then the contours ${\cal C}_a$ run along the real axis and do
not go through the logarithmic singularity. In this case, a change of
variables in the integral \fcftepa\ yields
\eqn\fordilog{\eqalign{{\cal L}_1(x)=&L(1/(1+x))\cr
{\cal L}_{-1}(x)=&-L(1/x),\cr}}
where
$$L(x)=-{1\over 2}\int_0^x dy\left[{\ln y\over (1-y)} + {\ln(1-y)\over
y}\right].$$
There are many known identities involving the Rogers dilogarithm function
$L(x)$, and they can often be used to evaluate \usingdilog\ exactly\rKR\rLewin.

Since this derivation allows an arbitrary chemical potential for each
particle, the expectation value of any symmetry operator diagonal on the space
of particles is obtained by setting $\lam_a$ to be the value of $A$ on the
particle $a$. A variety of such examples will be presented in sections 5 and
6.

\newsec{Non-diagonal operators}

In this section, we show how to calculate the expectation value of a symmetry
operator $K$ which is not diagonal on the space of particles. This involves
truncating the Hilbert space, just like in the (Ising$)^2$ model in the second
half of section 2.
The procedure for calculating
\eqn\forK{<K>\equiv \tr\left[ Ke^{-RH_L}\right]}
 basically follows that of section 3: we find the appropriate quantization of
momenta in periodic boundary conditions, and use this as a constraint while
minimizing the free energy. This results in an integral equation for the
pseudoenergy $\ep(\tht)$ and an expression for the free energy in terms of
$\ep(\tht)$. If there are no particles neutral under $K$, the asymptotics of
these equations are obtainable, allowing the calculation of the conformal
dimensions of the operator which creates the excited state. In others, it
seems that one must resort to a numerical calculation.

The relation which determines the level densities $\rho^a(\tht)$
remains \bigone, but we must modify
the free energy. The Hamiltonian is diagonal in theories with diagonal
S-matrices, so only states invariant under $K$ contribute to the sum in
\forK.\foot{We do not treat operators which take a particle into a
superposition of particle states.} Thus, as before, $<K>$ is the sum of
$e^{-RH_L}$ over states on which $K=1$.  This restriction changes the entropy.
The species of particles form multiplets under $K$, and all species of
particles in a multiplet must have the same level density and particle density
(this is true even in the unrestricted case). The crucial effect of the
restriction is that the entropy from only one species of each multiplet
contributes to the free energy. This is because in order to specify a given
state with $K=1$, one needs only to specify the set of momenta for one species
in a given multiplet; the other sets must be identical by $K$ invariance.
Minimizing the free energy, one finds the coupled integral equations are
modified to
\eqn\forepaii{\ep_a(\tht)=n_a m_a R\cosh\tht -\sum_b{1\over 2\pi}\int d\tht'
\phi_{ab}(\tht-\tht')\ln(1+\lam_b e^{-\ep_b(\tht)}),}
where $n_a$ is the number of species in the multiplet $a$.
The scaled free energy, defined as $f_K(mR)\equiv{R\over
L}\ln<K>_\lam=RE_K(R)$, is
\eqn\ffaii{\eqalign{f_K(mR)=&\sum_a {1\over n_a} f_\lam^a,\cr
f_\lam^a\equiv &-{mR\over 2\pi}\int d\tht \cosh\tht\ln(1+\lam_a
e^{-\ep_a(\tht)}).\cr}}

In the limit $mR\rightarrow 0$, doing the same substitutions as in in section
3, one finds that (specializing to the case where all $n_a$ are $1$ or $2$)
\eqn\ffaiii{f_K(0)= -c_\lam \left({\pi\over 6}\right)- {3\over 2} \sum_A f_A,}
where the sum over $A$ is only over particles with $n_A=2$, and $c_\lam$ is
the result obtained without truncating the Hilbert space; e.g., $c_1$ is the
central charge derived by the methods of section 3 with $\lam=1$.
Unfortunately, I can find a closed-form expression for $f_K(0)$ like \fcftepa\
only in the case where all the $n_a$ are the same. In the case where all
$n_a=2$, \ffaiii\ yields
\eqn\ffaiv{f_K(0)= -{c_\lam\over 4}\left({\pi\over 6}\right).}
When $\lam=1$, \BCN\ gives $h+\bar h=({3\over 4}c)/12 $, and the operator
which creates this excited state has conformal dimensions $(c/32,c/32)$. This
is quite an odd result---it applies to any theory with a conjugation symmetry
that leaves no particles neutral. This includes the situation where two
copies of a model are coupled by boundary conditions, as in the (Ising$)^2$
model treated in section 2.

\newsec{Examples from Toda minimal S-matrices}
\subsec{The Three-State Potts Model}

The three-state Potts model at its conformal point has central charge $c=4/5$.
Perturbation by the thermal operator leaves the theory integrable, and its
exact
S-matrix was conjectured in \rZamii. The particle spectrum consists of a
particle (labeled $1$) and its antiparticle $\bar 1$. These particles are in
the
two-dimensional representation of the model's $S_3$ symmetry. This symmetry is
generated by the operators $A$, which is diagonal and multiplies particle $1$
($\bar
1$) by $e^{2\pi i/3}$ ($e^{-2\pi i/3}$), and $K$, which exchanges the two
particles.

Calculating $<K>$ requires merely reading off the answer from section 4. The
conformal dimensions of the operator creating the associated excited state are
found from equation \ffaiv, and in this case are $(1/40,1/40)$. In the
language of the lattice Potts model (where a spin $\sigma$ takes the values
$1$, $e^{2\pi i/3}$ or $e^{-2\pi i/3}$), this corresponds to finding the
lowest eigenvalue of the Hamiltonian with ``twisted'' boundary conditions,
where $\sigma(0,y)=\sigma^*(R,y)$. This is in agreement with the conformal
result\rCarBC. We note that even though the (1/40,1/40) operator is not part
of the operator content of the three-state Potts model on the torus, it is
present with these boundary conditions. This is analogous to the situation in
the Ising model, where the free fermion is not part of the toroidal operator
content, but does belong in the case of antiperiodic boundary
conditions\rCarBC.

Calculating $<A>$ requires the introduction of a chemical potential, as in
section 3. In this case, we must have a complex fugacity: $\lam_1=e^{2\pi
i/3}$ and $\lam_{\bar 1}=e^{-2\pi i/3}$.  We define the TBA equations here by
analytic continuation from $\lam_1=\lam_{\bar 1}= 1$, whose TBA system was
derived in \rAlZi.  In this analytic continuation, we keep $\lam_1=(\lam_{\bar
1})^*$, enabling us to require that $\ep^1=(\ep^{\bar 1})^*$. This specifies
which branch of the logarithm is taken in \forepa\ and \ffa.
The free energy \ffa\ remains real
because $f^1_{\lam}=(f^{\bar 1}_{\lam^*})^*$.
These resulting TBA equations are those discussed in \rMar.

To evaluate $f$ in the conformal limit, we must find $x_a$ by solving \forxa.
In this case, $N_{11}=N_{22}=1/3$, and $N_{12}=N_{21}=2/3$.\rAlZi\ This
equation has more than one solution, but the correct one is the continuation
of the $\lam=1$ result of $x_1=x_{\bar 1}=(\sqrt{5} + 1)/2$. It is
$${x_1 \over e^{{2\over 3}\pi i}} ={x_{\bar 1}\over e^{-{2\over 3}\pi i}}=
{1\over 2}(1-\sqrt{5}),$$
which with our analytic continuation yields
\eqn\xpotts{\eqalign{x_1=&{1\over 2}(\sqrt{5} -1)e^{-i{\pi \over 3}}\cr
x_{\bar 1}=&{1\over 2}(\sqrt{5} -1)e^{+i{\pi \over 3}}.\cr }}
Thus the contour $C_1$ in \fcftepa\ starts at $\ln x_1=\ln[(\sqrt{5} - 1)/2] -
i\pi/3$ and goes off to positive infinity, asymptotically approaching the real
$\ep$-axis from below. It goes over the logarithmic singularity at
$\ep=i\pi/3$. The
contour $C_{\bar 1}$ is of course the complex conjugate of $C_1$. The actual
shape of these contours is determined by the solutions $\ep_a(\tht)$, but we
do not need this information to evaluate \fcftepa. We can deform the contour
$C_1$ so that $Im(\ep)$ is always $-i\pi/3$. This goes right through the
singularity of the integrand, but the integral is well-defined. Shifting
$\ep\rightarrow\ep+i\pi/3$ in \fcftepa\ yields
\eqn\fpotts{\eqalign{2f_A(0)=
-{1\over \pi}& \int_{\ln|x_1| }^{\infty} d\ep \left[{-\ep e^{-\ep}\over
1-e^{-\ep}}+\ln(1-e^{-\ep})+\left({i\pi\over 3}\right){e^{-\ep}\over
1-e^{-\ep}}\right]\cr
& + {\hbox{complex conjugate}}.\cr}}
The last term in the square brackets is a result of the shift, and can be
evaluated explicitly, giving
$$\left(-{1\over \pi}\right) {i\pi\over
3}\ln(1-e^{-\ep})\bigg|_{\ln|x_1|}^{\infty} + {\hbox{complex conjugate}}.$$
This is not zero because the argument of the logarithm at $\ep=\ln|x_1|$ is
negative, and the way we have analytically continued means that the logarithm
from $C_1$ gives $-i\pi$, whereas the logarithm from $C_{\bar 1}$ gives
$+i\pi$. Thus the total contribution of this term is $-2\pi/3$. To evaluate
the other two terms, we split the contour into two parts, one running from
$\ln|x_1|$ to zero, and the other from zero to infinity. Then the change of
variables $x=1- e^{\ep}$ in the first part and $x=e^{-\ep}$ in the second
yields
\eqn\fpottsii{ f_A(0)=-{1\over\pi}\left[ -2L({1\over 2}(3-\sqrt{5}))-2L(1)
+{\pi^2 \over 3}\right] .}
Since $L(1)=\pi^2/6$ and $L( {1\over 2}(3-\sqrt{5}))=\pi^2/15$, this yields
\eqn\Epotts{E_A=+{4\over 5}\piR.}
The operator which creates this state has dimensions $(1/15,1/15)$, as was
derived in \rMar. In the lattice Potts model, this corresponds to ``cyclic''
boundary conditions, where $\sigma(0,y)=\exp(2\pi i/3)
\sigma(R,y)$. This result has been derived at the conformal point\rCarBC.

\subsec{${\bf Z}_{n+1}$ parafermions}

The Ising model and the 3-state Potts model are the first two models in the
hierarchy of ${\bf Z}_{n+1}$ parafermions\rZF.  At the conformal point these
models have central charge $c=2n/(n+3)$. There are ``spin'' fields $\sigma_k$
($k=1\dots n$) which have conformal dimension $h=\bar h=k(n+1-k)/2(n+1)(n+3)$,
and ``C-disorder'' fields $\Phi_s$ ($s=0,1,2\dots \le (n+1)/2$), which have
conformal dimensions $h=\bar h=[n-1+(n+1-2s)^2]/16(n+3)$.
These conformal field theories remain integrable when perturbed by the primary
field with dimensions ($2/(n+3),2/(n+3)$), and the S-matrix is conjectured to
be the $A_{n}$ minimal Toda S-matrix\rKS.\foot{For arguments as to why this is
so, see \rEYHM.} There are $n$ particles with mass $m_a=\sin(\pi a/(n+1))$,
$a=1\dots n$.

The value of $N_{ab}$ for a simply-laced Toda minimal S-matrix are given by
the matrix relation $N=I(2-I)$, where $I$ is the incidence matrix ($\equiv 2 -
C$, where $C$ is the Cartan matrix) of the Lie algebra corresponding to that
Toda theory. It was shown in \rAlZv\ that the relation
\forxa\ for simply-laced Toda theories can be transformed to the simpler form
\eqn\forxaii{x_a=\prod_b (\lam_a + x_a)^{{1\over 2}I_{ab}}.}

The symmetry of a Toda S-matrix is the symmetry of the extended Dynkin
diagram. For $A_n$ this is the dihedral group ${\cal D}_{n+1}$, and is
generated by two operators $A$ and $K$. The diagonal ${\bf Z}_{n+1}$ symmetry
is generated by $A$, which when acting on the particle $a$ gives the phase
$\exp(2\pi ia/(n+1))$. $K$ implements a ${\bf Z}_2$ symmetry, which exchanges
particle $a$ with the particle $n+1-a$. For more details on these and other
Toda S-matrices, and on the TBA in these theories, see \rKMi\ and \rAlZv.

By using the method of section 3, we can determine $<A^k>$ by setting the
chemical potentials $\lam_a=\exp(i2\pi ka/(n+1))$.
For the $A_n$ Toda minimal S-matrices, the solution of \forxaii\ is
\eqn\xapara{{x_a\over\lam_a}={\sin{(k+1)a\pi\over n+3}
\sin{(k+1)(a+2)\pi\over n+3}\over\sin^2{(k+1)\pi\over n+3}}.}
These values should be used with \fcftepa\ to determine the ground-state
energy in the conformal limit. In the Ising ($n$=1) and three-state Potts
($n$=2) models, the excited state associted with $<A^k>$ is created by
$\sigma_k$.  Therefore, we conjecture that this holds true for this entire
hierarchy. This has been checked for $n=4$ in \rMar. In the next section, the
$n=3$ case is discussed in detail. However, I was unable to find a general way
of evaluating the integral in \fcftepa, and evaluating them case-by-case is
straightforward but tedious.

By using the method of section 4, we find the integral equations which
determine $<K>$. For $n$ even, there are no particles neutral under $K$, so we
can read off the answer in the conformal limit from \ffaiv. The operator which
creates the associated excited state has dimensions $n/16(n+3)$, and is the
$\Phi_{n/2}$ field. To evaluate $<K>$ in the conformal limit for
$n$ odd would require solving the integral equation \forepaii\ numerically,
and substituting the result into \ffaii. However, it can be easily estimated
using
\eqn\Kestimate{f_K(0)\approx\sum_a {1\over n_a^2}L({x_a\over 1+x_a})}
where $L$ is defined in \fordilog, and the $x_a$ are those of the untruncated
model. (The $x_a$ do not change because \forepa\ and \forepaii\ are identical
in the $m\rightarrow 0$ limit.) Evaluating the dilogarithms  numerically, we
find that $$f_K(0)\approx -{1\over 2}\left({\pi\over 6}\right)$$ for all odd
$n$.
The associated field has dimensions $(n-1)/16(n+3)$, and is $\Phi_{(n+1)/2}$.
One should be able to verify this in the conformal limit
by using the analysis of \rQiu.

\subsec{${\bf Z}_4$ parafermions (in detail)}

In this section we calculate the conformal limit of $<A>$ and $<A^2>$ for
${\bf Z}_4$ parafermions.  The S-matrix is that of the $A_3$ Toda theory,
where the mass spectrum consists of three particles $1$, $2$ and $3$, with
masses $1/\sqrt{2}$, $1$ and $1/\sqrt{2}$, respectively.

To calculate $<A^2>$, we set $\lam_2=1$, $\lam_1=\lam_3=-1$. The solution of
\forxaii\ is given in \xapara, and is $x_1=x_3=1$, and $x_2=0$. It is not the
unique solution, but it is easy to
verify that this in fact is the solution obtained by deforming $\lam_1$ and
$\lam_3$
continuously from $1$ to $-1$. Since all the $\lam_a$ are either
$1$ or $-1$, we use \usingdilog\ and \fordilog\ to obtain
\eqn\fzaii{\eqalign{f_{A^2}(0)&=-{1\over \pi}[L(1)
-2L(1)]\cr
&=\left({\pi\over 6}\right),\cr}}
since $L(1)=\pi^2/6$.
The equivalence \EAR\ means that this can be interpreted as the lowest
eigenvalue of the Hamiltonian with boundary conditions appropriate to $A$.
(The lattice-model interpretation of these boundary conditions will be given
in the next subsection.) The conformal dimensions of the operator creating
this state are given by the asymptotic formula \BCN, and are (1/12,1/12).
This operator is the spin field $\sigma_2$.

The calculation of $<A>$ in the conformal limit is a bit more involved. We
 set $\lam_a=(i)^a$,
and the solution of \forxaii\ is $x_1=0$, $x_2=1$. Using \fordilog, we find
$${\cal L}_{\lam_2}=-{1\over \pi}L(1)={\pi\over 6}.$$
Since $x_1$ and $x_3$ are real the contours in \fcftepa\ run along the real
axis:
\eqn\fza{{\cal L}_{\lam_1}+{\cal L}_{\lam_3}=-{1\over 2\pi}
\int_{-\infty}^{\infty} d\ep\left[{\ep ie^{-\ep} \over
1+ie^{-\ep}}+\ln(1+ ie^{-\ep})\right] + {\hbox{complex conjugate}}.}
Even though the two terms in the square brackets individually diverge when
$\ep\rightarrow\infty$, their sum does not, and the integral is well-defined.
Since the integral is over the real axis, we can add it and its complex
conjugate, obtaining
\eqn\fzaii{\eqalign{{\cal L}_{\lam_1}+{\cal L}_{\lam_3}&=-{1\over 2\pi}
\int_{-\infty}^{\infty} d\ep\left[{\ep  e^{-2\ep} \over
1+e^{-2\ep}}+\ln(1+ e^{-2\ep}) \right]\cr
&=-{1\over 2\pi}L(1),\cr}}
where we do the change of variables $x=1/(1+\exp(2\ep))$ to obtain the
dilogarithm in the second line. Adding all the contributions together, we
obtain
\eqn\fzaiii{f_A(0)={1\over 2}\left({\pi\over 6}\right),}
and associated operator is the spin field
$\sigma_1$, with conformal dimensions of (1/16,1/16). We also have
$<A^3>=<A>$, which agrees with the general parafermion symmetry
$\sigma_k\leftrightarrow\sigma_{n+1-k}$.

\subsec{$D_n$ Toda theories}
The $D_n$ Toda S-matrices describe a hierarchy of models which in the
conformal limit all have $c=1$. They correspond to a scalar field with radius
$\sqrt{n/2}$ orbifolded by its ${\bf Z}_2$ symmetry \rGin. They are the
continuum limit of the lattice Ashkin-Teller model, which consists of two Ising
spins at each site with a four-spin coupling specified by the value of $n$.
The (Ising$)^2$ model and the ${\bf Z}_4$ parafermions are the first two
models in this series, since $D_2=A_1+A_1$ and $D_3=A_3$. The $D_n$ model has
$n$ particles, with $m_0=m_{\bar 0}=1$ and $m_a=2\sin(\pi a/2(n-1))$, where
$a$ runs from $1$ to $(n-2)$. As before, the symmetry is that of the extended
Dynkin diagram. For more details of these models, see \rBCDS\ or
\rKMi. The TBA was applied to these models in \rKMi.

The symmetry structure is slightly different for $n$ odd or even. For all $n$,
there is a ${\bf Z}_2$ generated by $K$, which interchanges $0$ with $\bar 0$.
(For n=4, this becomes $S_3$, which interchanges $1,0$ and $\bar 0$.)  For $n$
even, the diagonal symmetry is ${\bf Z}_2\times{\bf Z}_2$, and is generated by
two operators $(A_1,A_2)$, which act on the particles
 with eigenvalues $\alpha_1$ and $\alpha_2$, where
$\alpha_1(a)=\alpha_2(a)=(-1)^a$, $\alpha_1(0)=\alpha_2(\bar 0)=1$, and
$\alpha_1(\bar 0)=\alpha_2(0)=-1$.  For $n$ odd, the diagonal symmetry is
${\bf Z}_4$, and is generated by $A$, with eigenvalues $\alpha(a)=(-1)^a$,
$\alpha(0)=i$ and $\alpha(\bar 0)=-i$.

To calculate $<A_1A_2>$ for $n$ even or $<(A)^2>$ for $n$ odd, we set
$\lam_0=\lam_{\bar 0}=-1$ and $\lam_a=1$.  The solution of \forxa\ is
\eqn\xadi{\eqalign{x_0=x_{\bar 0}&=1\cr
 x_a&={\sin({(a-1)\pi\over n})\sin({(a+1)\pi\over n})\over\sin^2({\pi\over
n})}.\cr}}
Since all $\lam_a$ are $\pm 1$, \usingdilog\ and \fordilog\ are applicable in
the conformal limit.  Notice that $x_1=0$ and that the $x_a$ for $a\ge 2$ are
those for the $A_{n-3}$ Toda theory, as displayed in \xapara\ with $k=0$. Thus
\eqn\fdaa{\eqalign{f_{(A_1A_2)}(0)=&-{\pi\over 6}({2(n-3)\over n} + 1 -1 -1)\cr
=&-{\pi\over 6}(1-{6\over n}).\cr}}
The associated excited state is created by the field with conformal dimensions
($1/4n,1/4n$). This corresponds to anti-periodic
boundary conditions on both Ashkin-Teller spins, and has been derived in the
conformal limit\rYang.

To calculate $<A_1>=<A_2>$ for $n$ even, we set $\lam_0=1$, $\lam_{\bar 0}=-1$
and
$\lam_a=(-1)^a$. The solution of \forxa\ is
\eqn\xadii{\eqalign{x_0=x_{\bar 0}=&1\cr
x_a=&1 \qquad\qquad a\quad {\hbox{odd}}\cr
x_a=&0 \qquad\qquad a\quad {\hbox{even}}.\cr}}
Thus \usingdilog\ and \fordilog\ yield
\eqn\fdaaii{\eqalign{f_{A_1}(0)=&-{\pi\over 6}(-1+{1\over 2} + 1 -1+\dots +1
-1)\cr
=&{1\over 2}\left({\pi\over 6}\right).\cr}}
The associated excited state is created by the field with dimensions
($1/16,1/16$) and is called the spin field. This corresponds to anti-periodic
boundary conditions on one of the Ashkin-Teller spins, and has also been
derived in the conformal limit\rYang.

Calculating $<A>$ for $n$ odd requires setting $\lam_0=i$, $\lam_{\bar 0}=-i$
and
$\lam_a=(-1)^a$. The solution of \forxa\ is \xadii\ in this case as well, but
we cannot use \usingdilog\ this time. However, evaluating \fcftepa\ for the
three particles $0,\bar 0$ and $1$ is identical to the calculation of $<A>$
done for the $D_3$ model (${\bf Z}_4$ parafermions) treated in the previous
subsection. The rest of the terms in \fcftepa\ can be evaluated with
\usingdilog, so we find
\eqn\fdaaiii{\eqalign{f_{A}(0)=&-{\pi\over 6}(-{1\over 2} + 1 -1+\dots +1
-1)\cr
=&{1\over 2}\left({\pi\over 6}\right),\cr}}
and the associated field has conformal dimensions ($1/16,1/16$) here as well.

Evaluating $<K>$ exactly for $n> 2$ would require a numerical calculation.
However, using the approximate formula \Kestimate\ yields conformal dimensions
of
\eqn\KK{(h,\bar h)\approx \left({1\over 16}L(1/n),{1\over 16}
L(1/n)\right) \approx({1\over 16n},{1\over 16n}),}
which is the result obtained at the conformal point.

\subsec{$E_6, E_7$ Toda theories}

The tricritical Ising model ($c=7/10$) perturbed by its thermal operator is
believed to be described by the seven-particle $E_7$ Toda S-matrix\rCMFKM.
The solution of
\forxaii\ for $\lam_a=1$ is given in \rKMi.  This model has a diagonal ${\bf
Z}_2$ symmetry which multiplies particles $1$, $3$ and $6$ (in the notation of
\rKMi) by $-1$. The solution of \forxaii\ for $\lam_1=\lam_3=\lam_6=-1$ is
\eqn\forevii{\eqalign{x_4=x_7=&0\cr x_1=x_3=x_6=&1\cr
x_2=x_5=&{1+\sqrt{5}\over 2}.\cr}}
Thus \usingdilog\ and \fordilog\ yield
\eqn\evii{\eqalign{f_A(0)=&-{1\over 2\pi}\left(-1+{2\over 5}-1+1+{2\over
5}-1+1\right)\cr
=&{1\over 5}\left({\pi\over 6}\right),\cr}}
 which means that the associated operator has conformal dimensions
(3/80,3/80).

The tricritical three-state Potts model ($c=6/7$) perturbed by its thermal
operator is believed to be described by the six-particle $E_6$ Toda
S-matrix\rFZSZ.  The solution of \forxaii\ for $\lam_a=1$ is given in \rKMi.
This model has an $S_3$ symmetry generated by the operators $A$ and $K$. The
diagonal operator $A$ multiplies particles $1$ and $3$ by $\exp(2\pi i/3)$,
$\bar 1$ and $\bar 3$ by $\exp(-2\pi i/3)$, and leaves particles $2$ and $4$
neutral (using the labels of \rKMi). $K$ interchanges $1$ with $\bar 1$ and
$3$ with $\bar 3$, while leaving $2$ and $4$ neutral. The calculation of
$f_A(0)$ was outlined in \rMar, and the associated operator has dimensions
$(1/21,1/21)$. To calculate $f_K(0)$ would require a numerical calculation,
since there are particles left neutral by this symmetry. The approximate
formula \Kestimate\ yields the conformal dimensions of (1/56,1/56) for the
associated operator.

\newsec{Examples from the minimal models}

\subsec{The massless perturbation of the tricritical Ising model}

The conformal tricritical Ising model (the $n=4$ minimal model) perturbed
by the $\phi_{1,3}$ operator
$$S_{TCI}\longrightarrow S_{TCI} + g\int d^2x \phi_{1,3}$$
is described by two different field theories, depending on the
sign of $g$.  For $g$ positive, one obtains a theory of massive kinks with a
non-diagonal S-matrix\rZamkink. Since the S-matrix is not diagonal, one must
introduce pseudoparticles into the TBA\rAlZii. Due to the difficulty in
undersanding the effect of boundary conditions on the pseudoparticles, we will
not discuss this case here. When $g$ is positive, the model flows to the Ising
model ($n=3$ minimal model). Thus there must be massless excitations, which
are the only particles which will remain in the infrared limit. As shown in
\rAlZiii, one would expect there to be a right-moving and a left-moving
massless fermion, which in the infrared limit become the free Majorana fermion
of the Ising model. The S-matrix for these fermions was conjectured in
\rAlZiii, and the TBA system was derived. Since these particles are massless,
the rapidity relations \rap\ do not apply but instead are replaced by
\eqn\rapTCI{\eqalign{&P_{right}={M\over 2} e^\tht\cr
&P_{left}=-{M\over 2} e^{-\tht},\cr}}
where $M$ is the scale of the theory, which depends on $g$. The TBA equations
\forepa\ are changed accordingly.

We calculated $<(-1)^N>$ in the Ising model in section 2, and showed that it
corresponded to periodic boundary conditions in the $R$-direction on the
fermion. The same is true here. The calculation can be done by introducing
fugacities $\lam_{right}=\lam_{left}=-1$. The analysis of section 3 is easily
repeated to take into account the modified rapidities\rapTCI. One then obtains
the TBA equations conjectured in \rMar\ and \rKMiv\ as describing the
excited-state energy for this model. We thus have provided a physical
interpretation of these equations. These papers show that the operator which
creates the excited state has dimensions (3/80,3/80) in the conformal limit.

\subsec{The ``staircase'' model}

A simple TBA system with many intriguing similarities to the minimal models
has recently been studied\rAlZiv. The system consists of a single particle
$\Phi$ with mass $m$ and with S-matrix
\eqn\Smatphi{S(\tht)={\sinh\tht-i\cosh\tht_0 \over \sinh\tht+i\cosh\tht_0},}
where $\tht_0$ is a free parameter. Using \forphi\ and \fornab, one finds
$N_{\Phi\Phi}=-1$. Solving \forxa\ yields $x_\Phi= 0$, which means that in
the $m\rightarrow 0$ limit, $c=1$. This S-matrix is the analytic continuation
of the S-matrix for the sinh-Gordon model\rssG
\eqn\ssg{S_{shG}=\int \left[{1\over 2}(\del_\mu\Phi)^2 -
2\mu\cosh\beta\Phi\right] d^2 x.}
to complex coupling $\gamma={\pi\over 2}+i\tht_0$, where
$$\gamma={\beta^2/8\over 1 + \beta^2/8\pi}.$$

In \rAlZiv\ it is shown that at large $\tht_0$, $f(mR)$ develops an unusual
``staircase'' structure. For all other known models, the structure of $f(mR)$
(as a function of $mR$ on a log scale) is a plateau at $f(0)$, which
eventually falls off smoothly to $f(\infty)=0$. One then can read off the
central charge of the conformal theory from the value of $f(0)$. In this
``staircase'' model at large $\tht_0$, $f(mR)$ develops plateaus in various
regimes of $mR$, making the plot of $f(mR)$ look something like a
staircase. Even more intriguing is that the values of $f(mR)$ on the plateaus
are the values which give the central charge of the minimal models,
$c=1-6/n(n+1)$. Thus the picture one has is that one starts out at a $c=1$
conformal theory at $m=0$, and then as $m$ is increased, the renormalization
group flow takes one into the neighborhood of the fixed point corresponding to
each of the minimal models. The trajectory spends some time in the
neighborhood of each model, before the instability takes over, forcing it to
flow to the next model. This goes on until it finally reaches the $c=0$
trivial model, where all the degrees of freedom have been scaled away. The
staircase structure was derived numerically and analytically. It is derived
analytically by showing that the TBA system defined by \Smatphi\ can be
accurately approximated by the TBA system conjectured in \rAlZii\ as describing
the flow of the ground state from the $n$-th to the $(n-1)$-th minimal
model\rLC. For a much more detailed description of the staircase model, see
the original paper\rAlZiv.

Since there is no pole in the S-matrix \Smatphi\ corresponding to $\Phi$ as a
bound state of itself, this model must have a ${\bf Z}_2$ symmetry
$\Phi\rightarrow -\Phi$. This is also apparent from the sinh-Gordon
description \ssg. Thus in our usual manner we can set $\lam=-1$ and calculate
$<(-1)^N>$. The solution of \forxa\ is $x_{\Phi}=2$, so
\eqn\fstair{f_{-1}(0)={1\over 2}({\pi\over 6}).}
Thus the associated operator in the conformal limit has dimensions
($1/16,1/16$). It is easy to repeat the argument of \rAlZiv\ to show that
$f_1(mR)$ develops plateaus, only this time the TBA system can be approximated
by the TBA system studied in sections 5.1 and 5.3 of ref.\rKMiv. This is
identical to the system conjectured for the ground-state energies\rAlZii,
except in this case all the $\lam_a=-1$ instead of $1$. The plateaus do not
form a staircase here; for $n$ odd (even) the plateaus are at values greater
(less) than $f_{-1}(0)$. The plateaus take values of $f_{-1}$ corresponding to
the conformal dimensions of the field $\phi_{n/2,n/2}$ for $n$ even and
$\phi_{(n+1)/2,(n+1)/2}$ for $n$ odd. This is conjectured to describe the flow
from the $n$-th to the $(n-1)$-th minimal model of an excited-state energy for
$n$ even, but not $n$ odd\rKMiv. This is for several reasons. The first is
that the numerical results for $n$ odd beyond lowest order do not agree with
perturbation theory around the conformal point. The second is that one expects
the flows in minimal models to take the $\phi_{i,i}$ operator in the $n$-th
minimal model to the $\phi_{i,i}$ operator in the $(n-1)$-th minimal
model\rZamLG\rLC; this is true here only for $n$ even.

So while our result shows that excited states in the staircase model resemble
those in the minimal models in some respects, there is a crucial difference.
Although the flow down the chain of operators seems very natural from the
point-of-view of the staircase model, its relation to flows in the minimal
models is still not completely clear.

\newsec{Conclusions and speculations}

We have seen that the TBA is useful for calculating not only the Casimir
energy of the lowest-energy state in the theory, but for a number of excited
states as well. This is done by choosing the boundary conditions to project
out the ground state. In the limit $R\rightarrow\infty$, all of the states
discussed here have $E\rightarrow 0$, and become degenerate with the ground
state.\foot{I thank Tim Klassen for pointing this out.} It is not clear if it
will be possible to project out all such states and obtain information on
massive states, in the manner of \rKMiii\rLM.

The most obvious open direction is to understand how these results can be used
to study excited states in theories with non-diagonal S-matrices, where one
must introduce massless pseudo-particles into the TBA\rAlZii. TBA equations in
some such situations have been conjectured, but it is not obvious how to
interpret these equations as boundary conditions on the field theory. However,
this hopefully is only an interpretational issue: all the formalism should be
applicable to these cases.

There are many other operator expectation values that one can calculate. For
example, it is straightforward to calculate $<(-1)^{N_a}>$ for any particle
$a$ in a diagonal scattering theory, since $(-1)^{N_a}$ is a symmetry of the
S-matrix even if it is not a symmetry of the field theory or of the fusion
rules. The interpretation of this in terms of boundary conditions is not
immediately obvious. However, in the minimal models on the lattice (on and off
criticality) it is possible to define boundary conditions so that any of the
primary operators creates the lowest-energy state\rDJMO.  This seems to be
true for all rational conformal field theories as well\rCarVer. It is thus not
difficult to believe that once suitable ways are found of implementing the
boundary conditions, one will be able to calculate the energy for all such
excited states by calculating operator expectation values with the TBA.

\bigskip
\centerline{\bf Acknowledgements}

I would like to thank Ken Intriligator, Tim Klassen, Kolya Reshetikhin and
Cumrun Vafa for helpful and interesting conversations. I would also like to
thank Michael L\"assig for mailing me various papers.
This work was supported
by Department of Energy grant DEAC02-89ER-40509, and by National Science
Foundation grant PHY-905-7173.

\vfill
\eject
\listrefs
\bye\end